# Methods and Tools for the Management of Renewable Energy Communities: the ComER project

Anna Rita Di Fazio, Arturo Losi, Mario Russo
*Department of Electrical and Information Engineering*
*University of Cassino*
Cassino, Italy
a.difazio@unicas.it

Filippo Cacace, Francesco Conte, Giulio Iannello
*Departement of Engineering*
*Campus Bio-Medico University of Rome*
Rome, Italy
f.conte@unicampus.it

Gianluca Natrella, Matteo Saviozzi
*Departmenet of Electrical, Electronics, Telecommunications and Naval Architecture*
*University of Genova*
Genova, Italy
gianluca.natrella@edu.unige.it

*Abstract*— Renewable Energy Communities (RECs) have been officially introduced into the European legislation through the Clean Energy for all Europeans package. A REC is defined as an association of citizens, commercial activities, enterprises, and local authorities that own small-scale power plants based on Renewable Energy Sources (RESs). The community has the objective of maximizing the share of renewable energy, *i.e.* the self-consumption of the energy generated by the community RES power plants and to generally optimize the use of electrical energy. This paper describes the ComER project, developed by the University of Cassino and the Campus Bio-Medico University of Rome. The project focuses on the main technical problems to face for the realization of a REC. The principal objective is to develop methods and tools necessary for the management and control of RECs. In particular, this paper describes the rules established for RECs in the Italian legislations, the organization of the ComER project, the adopted solutions and the first obtained results.

*Keywords—Energy Communities, Renewables, Self-Consumption Management and Control.*

## I. Introduction

In the current decade, a significant deployment of Distributed Energy Resources (DERs) is expected with the aim of fostering the widespread exploitation of Renewable Energy Sources (RESs) and increasing the efficiency of energy use. Differently from large traditional power plants, DERs are high in number, connected to the MV/LV networks, and usually installed together with commercial and residential loads. The ratio between the power capacity installed annually through new DERs and the one from new centralized power plants is estimated to reach 5:1 by 2024 [1]. It is moreover estimated that the revenue of residential DERs market could more than quadruple within ten years.

This means that an evolution from a centralized energy production model to a mixed one is occurring, and an increasing role is played by DERs from RESs. This mixed model makes it possible to overcome the technical, economic and, in some cases, regulatory barriers that limit the development of RESs. The latter, in fact, still have higher costs than centralized production from fossil fuels and, above certain capacity levels, they have a heavy impact on the operation of both distribution and transmission systems, making more complex their management and control. The mixed model, on the other hand, promotes the self-consumption of energy produced by RESs and a more efficient end-use of energy.

The complexity of the processes leading to the massive deployment of DERs requires the involvement and sharing of objectives with all actors at different decision-making levels (international, national, regional and local). It also requires a better definition of the roles of some actors and the introduction of new ones. For example, countries that have set the most ambitious targets in terms of reducing $CO_2$ emissions, such as Denmark and Germany, have defined tailored DERs' development policies based on the needs of end-users and involving municipalities, stakeholders, local authorities, common citizens and their associations in the process [2],[3]. It can be said that the large-scale deployment of DERs requires changing the user paradigm from passive to active for two reasons: from a technical point of view, because the user becomes a *prosumer*, and from a decision-making point of view, because the user is directly involved in financing new technical solutions to pursue common objectives, both individual (such as the economic ones) and collective (such as the environmental ones).

In this framework, international energy policies identify Renewable Energy Communities (RECs) as key instruments for the promotion of DERs [4][5]. The European Union (EU) introduced RECs with Regulation 2018/2001/EU [6], which is part of the Clean Energy for all Europeans Package, whose aim is the development of a sustainable, competitive, secure and decarbonised energy system through the reduction of greenhouse gas emissions by 40% by 2030. Specifically, the Regulation introduces the possibility for citizens, companies and local authorities of the EU states to produce, consume, store and sell energy obtained from RESs by installing ESSs without paying charges, fees or taxes. The scheme refers to small-scale power plants (initially below 25 kW, now below 1 MW). The Regulation also defines possible forms of aggregation (companies, associations, foundations, cooperatives) and support policies (subsidised financing, awareness campaigns on economic and environmental benefits, etc.). Within a few years, RECs could revolutionise the energy market, leading to an energy model that is no longer centralized, but distributed throughout the territory, in which most of users are prosumers. By 2050, half of Europe's citizens could not only produce, but also manage their own energy. In some studies, RECs are estimated to reach the 45% of total EU demand by 2050.

In Italy, between 2020 and 2021, the EU Regulation has been transposed into law (law 8/2020 and D.lgs. RED II Nov. 2021). RECs are defined as associations between citizens, commercial activities, enterprises and/or local authorities that decide to join forces to equip themselves with power plants

Funded from POR FESR Lazio 2014-2020 under project no. A0375-2020-36770.

for producing and sharing energy from RESs. The objective of a REC is to generally optimize the use of energy and, specifically, to maximize the share of renewable energy. To carry out such an optimal operation, the REC should implement suitably designed methodological and technological tools. In this paper, we describe the ComER project ("Methods and tools for managing renewable energy communities"), developed by the University of Cassino (Italy) and the Campus Bio-Medico University of Rome (Italy). The project has the objective of developing methods and technological tools to optimally manage a REC according to the Italian rules. In the following, we will introduce the organization of the ComER project, the considered REC configurations, the adopted management and control methods and some examples of the obtained results.

The reminder of the paper is organized as follows: Section II summarizes the rules introduced for the RECs by the Italian legislation; Section III describes the ComER project objectives and its organization; Section IV introduces the study case REC configurations; V shows the first results of the project; finally, Section VI provides the conclusions of the paper.

## II. RECs in the Italian Legislation

As mentioned in Section I, in Italy RECs are defined by laws 8/2020 [7] and D.lgs. RED II Nov. 2021. The principal rules are summarized in the following.

- A REC is based on open and voluntary participation, is autonomous and effectively controlled by its members that constitute a legal entity.
- No REC member may engage in the sale of energy as a principal business.
- The REC must be equipped with RES power plants, owned by the community or by some of the members.
- The community can be equipped with ESSs.
- All REC components (loads, RES power plants and ESSs) must be connected under the same primary HV/MV substation.

The connection among the REC members is virtual: they are conventionally connected to the public grid, but they must be equipped with smart meters which makes available the measurements of their energy consumption and of the generated renewable energy during each hour of the day. Based on these measurements, the REC receives a remuneration consisting of three components:

i. the restitution of a part of the energy use tariff, according to a cost-reflective network charge remuneration logic;
ii. an incentive proportional to the *shared energy*;
iii. the remuneration of the energy delivered to the grid by each RES power plant at the market price.

The *shared energy* is defined as the portion of the renewable energy delivered to the grid during a given hour, consumed in the same hour by the community members. Mathematically, it is defined as the minimum between the delivered renewable energy and the aggregated consumption of the REC members.

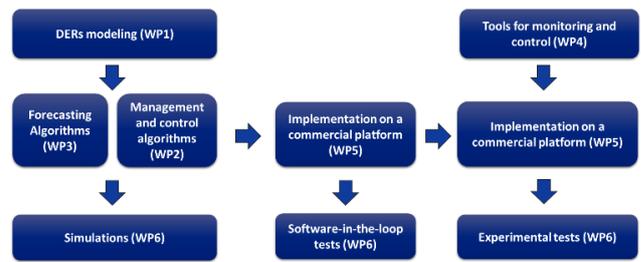

Fig. 1. Flowchart of the ComER Project.

Two types of RECs can be established: the "Communities of Collective Self-Consumption" (in Italian "Comunità di Autoconsumo Collettivo" (AUC)) and the effective "Communities of Renewable Energy" (in Italian "Comunità di Energia Rinnovabile" (CER)). The differences between these two types of RECs are that in the AUC the members must be in the same building or condominium and the RES power plant must be just one and with a maximal nominal power of 200 kW, whereas the CER members can be in different buildings, always under the same primary HV/MV substation, and the RES power plants can be more than one and with a maximal nominal power of 1 MW.

This implies a difference in the first component of the remuneration. Indeed, since in the case of AUCs the shared energy is physically self-consumed by the community, the distribution network charge is reduced more than in the case of CERs, where the shared energy is physically injected to the MV/LV grid. Therefore, the restitution of the energy use tariff is higher for AUCs (estimated to be about 10 €/MWh) and lower for CERs (estimated to be around 8 €/MWh). There is a difference also in the incentive paid for shared energy, which is equal to 100 €/MWh for AUCs and 110 €/MWh for CERs.

In the following we will indicate as RECs both AUCs and CERs, when no specific differentiation is required.

## III. The ComER Project

The overall objective of the ComER project is to develop methods and tools necessary for the management and control of RECs. Such methods and tools are certainly not new in the energy sector. In this case, however, they should be extended and adapted to take into account the specific features of RECs. Indeed, a REC will be composed of several user installations, as well as of RES power plants, controllable and not controllable electrical loads and ESSs, that may fall under the ownership and responsibility of different actors. We are therefore faced with a management and control problem where different actors have individual objectives and share common goals, which requires innovative approaches, methods and tools.

Fig. 1 shows the flowchart of project. It is organized into six Work Packages (WPs) shortly described in the following.

1. *DERs modeling*: suitable models representing the dynamics of DERs potentially installed within a REC will be defined; the detail level of models should be enough to represent their constraints in terms of power and/or energy capacity, controllability, and flexibility. The potential considered DERs components will be: Photovoltaic (PV) power plants, Battery Energy Storage Systems (BESSs), electrical vehicles (EVs) recharging stations, and flexible

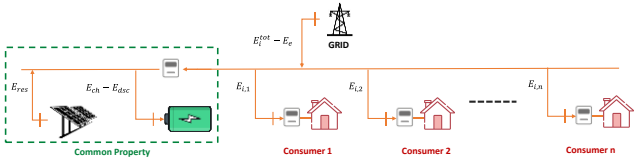

Fig. 2. REC Configuration 1.

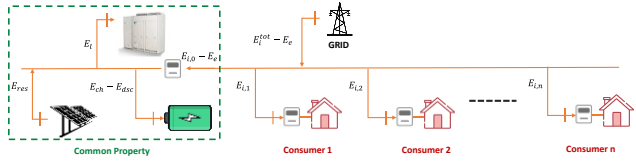

Fig. 3. REC Configuration 2.

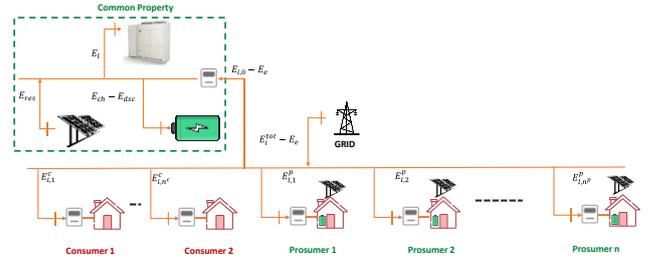

Fig. 4. REC Configurtion 3.

## A. REC configurations

RECs can assume different architectures, depending on the presence of the resources owned by the single members and the ones shared by the community. RES power plants and BESSs, for instance, could be installed on single residential loads (thus owned by single members) or on a public property (thus owned by the community). The community members, for their part, can be consumers or prosumers, and passive or active. Passive members are not willing to change their energy consumption profile, whereas active members will do so if a reward is granted. Moreover, the amount of flexibility provided by active members will depend on their personal availability to change the daily schedule of use of electrical energy, but also on the availability of devices such as smart household appliance and/or BESSs.

Figs. 2-4 show the three possible REC configurations considered in the ComER project. In Fig. 2 the REC Configuration 1 is presented: here the common property is a PV plant coupled with a BESS and the members are all consumers. They can either be passive or active. Fig. 3 shows another REC configuration where the common property also includes a load and the members are all consumers. Even in this configuration, the members can either be active or passive. The last configuration is shown in Fig. 4: here the common property is equal to the one of Configuration 2 but at least a set of members are prosumers as they possess their own PV panels and a BESS.

It is worth remarking that, potentially, the RES power plants can be other than PV systems (*e.g.* wind turbines, micro hydro generators, etc.) and ESS can be different form batteries (*e.g.* electrolyzer-fuel cells, reversible fuel cells, etc.). However, the ComER project focuses on PVs an BESSs since they are going to be the predominant technologies in Italy.

## B. Problem formulation

According to the Italian legislation, the community obtains a common income equal to:

$$I(k) = (t_r + t_{mise})E_s(k) + p_e(k)E_e(k) - t_i(k)E_{i,0}(k) \quad (1)$$

where: $k$ is the hour, $E_s$ is the shared energy, $E_e$ is the exported RES energy, $E_{i,0}$ is the energy imported by the common property, $t_r$ and $t_{mise}$ are the incentives bestowed for the restitution of tariff components and for the shared energy, respectively, $p_e$ is the zonal electricity price paid for the exported energy, and $t_i$ is the tariff paid for the energy import of the common property. The definitions of $E_s$ and $E_e$ depend on the REC configuration, whereas $E_{i,0}$ is present only if the common property has an internal load, which occurs in Configurations 2 and 3.

loads, such as the buildings air temperature regulation systems.

2. *Methods for RECs management and control*: the management and control methods should be able to combine the pursuit of the individual actors' own objectives (*i.e.* minimizing the electrical energy consumption) and the common objectives of the REC (*i.e.* maximizing the shared energy); both centralized and distributed approaches will be adopted. As a centralized approach, deterministic, stochastic, and scenario-based Model Predictive Control (MPC) methods will be studied; as a distributed approach, stochastic distributed optimization and multi-agent models will be considered.

3. *Forecasting algorithms*: suitably forecasting algorithms will be developed for PV generation and for loads; both conventional forecasting methods and ad hoc feature extraction or deep learning networks will be evaluated.

4. *Tools for monitoring and control*: the platforms available to implement advanced monitoring, prediction and control functions of a REC will be studied. A hardware/software architecture for collecting and processing data and sending signals for REC monitoring and control will be identified.

5. *Implementation on a commercial platform*: all the monitoring, management and control functionalities will be implemented in a commercial platform, obtaining, in this way, a prototype tool.

6. *Validation*: the prototype tool will be validated through synthetic simulations, Software-in-The-Loop simulations and experimental tests carried out on a real study case.

## IV. CONFIGURATIONS AND PROBLEM FORMULATION

In this section, we first introduce the REC configuration considered in the ComER Project and then we provide the formulation of the optimization problem to be solved in each of these configurations.

In Configuration 1, we have that:

$$E_c = \sum_{j=1}^{n} E_{i,j} \quad (2)$$

$$E_e = E_{res} + E_{dsc} - E_{ch} \quad (3)$$

$$E_s = \min(E_e, E_c) \quad (4)$$

where: $E_{i,j}$ is the energy demand of the $j$-th member, $E_c$ is the members aggregated energy demand, $n$ is the number of the members in the REC, $E_{res}$ is the energy generated by the PV power plant, $E_{dsc}$ and $E_{ch}$ are the energies exported and imported by the BESS, respectively. Notice that in (4) the shared energy $E_s$ is defined according to the Italian law definition, *i.e.* the minimum between the delivered renewable energy and the aggregated consumption of REC members.

The BESS discharge and charge energies $E_{dsc}$ and $E_{ch}$ must satisfy the following constraints:

$$0 \le E_{ch} \le \delta_b E_b^{nom} \quad (5)$$

$$0 \le E_{dsc} \le (1 - \delta_b) E_b^{nom} \quad (6)$$

where $\delta_b$ is a binary variable and $E_b^{nom}$ is the maximal energy that the BESS can exchange within one hour.

In Configuration 2, $E_c$ is given by (2) and $E_s$ is given by (4). The exported energy satisfies the following constraints:

$$E_e - E_{i,0} = E_{res} + E_{dsc} - E_{ch} - E_l \quad (7)$$

$$0 \le E_e \le \delta_e E_g^{nom} \quad (8)$$

$$0 \le E_{i,0} \le (1 - \delta_e) E_g^{nom} \quad (9)$$

where: $E_l$ is the energy demand of the internal load of the common property, $\delta_e$ is a binary variable and $E_g^{nom}$ is the maximal energy that the grid can exchange within one hour.

In Configuration 3, we have $n^c$ consumers and $n^p$ prosumers. The $j$-th consumer imports the energy $E_{i,j}^c$, which can be only a positive quantity. The $j$-th prosumer exchanges with the grid the energy $E_{l,j}^p$, which is positive when its RES generation is lower than its energy demand and negative in the opposite situation. In this second case, the prosumer is exporting renewable energy. Therefore, $E_c$ and $E_e$ are defined as follows:

$$E_c = \sum_{j=1}^{n^c} E_{i,j}^c + \sum_{j=1}^{n^p} \max(0, E_{l,j}^p) \quad (10)$$

$$E_e - E_{i,0} = E_{res} + E_{dsc} - E_{ch} - E_l + \sum_{j=1}^{n^p} \max(0, -E_{l,j}^p) \quad (11)$$

whereas the shared energy is always given by (4).

## V. First Results

In this section we present the first results of the ComER project. In particular, we show some outputs of WP3 and WP2 focused on the development of the forecasting algorithms and of the management and control methods, respectively.

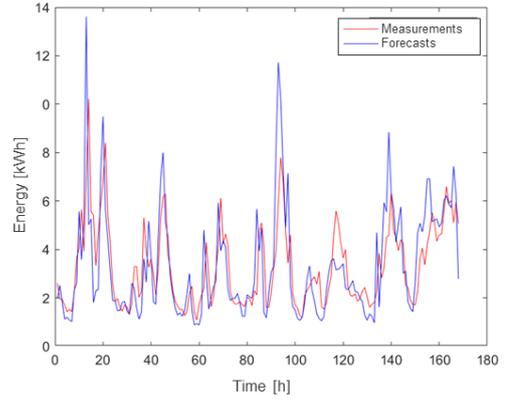

Fig. 5. Example of one-hour ahead forecast for an aggregate of residential loads with total nominal power of 42 kW.

### A. Forecasting alghorithms

The time series to be forecasted in order to optimally operate a REC are the ones of RES generation and of loads energy consumption. The first phase of the project acivities has been focused on the forecast of load consumptions. Within our scheme, the REC management agorithm needs, at least, a forecast of the aggregated energy consumption of REC members and of eventual common properties, with a granularity of one hour. The considered methods are based on autoregressive (AR) and autoregressive integrated moving average (ARIMA) methods. The objective is to obtain three types of forecasts: day-ahead, computed at a ginven hour of the day before for the 24 hours of the day after; one-hour-ahead, computed at each hour just for the following one; and 24-hours-ahead, computed at each hour for the subsequent 24 hours.

Fig. 5 shows an example of the obtained results for a one-hour-ahead forecast adopting multiple linear regressions with exogenous and endogenous variables, applied to an aggregate of residential loads with a total nominal power of 42 kW.

### B. Methods for REC management and control

In this paper we propose a solution developed for the REC Configuration 2 with passive consumers. The method adopts a deterministic MPC [10] to maximize the income of the REC. The MPC technique consists in solving an optimization problem at every discrete time $k$ over a finite horizon of length $T$, obtaining an optimal control trajectory $\{u^*(\tau)\}_{\tau=k}^{k+T-1}$. Then, according to the *receding horizon principle*, the first step of the optimal control trajectory $u^*(k)$ is applied and the same procedure is repeated at the following discrete time step. The advantages of MPC are well known. First of all, the receding horizon principle allows obtaining a closed-loop control. Indeed, at each time step, the optimization is repeated according to the new measurements of the system state and, in this way, modelling and forecasting errors can be compensated. Moreover, differently from other control methods, MPC allows the inclusion of state and control constraints, which are useful in practical applications.

In Configuration 2, the control variables are $E_{dsc}$ and $E_{ch}$. Therefore,

$$\{u(\tau)\} = \{E_{ch}(\tau), E_{dsc}(\tau)\} \quad (12)$$

and the MPC optimization problem is formulated as it follows:

TABLE I. SIMULATION PARAMETERS

| $C_b$ [kWh] | $S^{max}$ [p.u.] | $S^{min}$ [p.u.] | $\eta_{ch}$ - | $\eta_{dsc}$ - | $E_b^{nom}$ [kWh] | $E_g^{max}$ [kWh] | $t_r$ [€/MWh] | $t_{mise}$ [€/MWh] | $T$ [h] |
|---|---|---|---|---|---|---|---|---|---|
| 250 | 0.95 | 0.05 | 0.95 | 0.95 | 60 | 100 | 8 | 110 | 24 |

$$\{u^*(\tau)\}_{\tau=k}^{k+T-1} = \arg\max\left\{\sum_{\tau=k}^{k+T-1} I(\tau)\right\} \quad (13)$$

s.t. for all $\tau = k, k+1, \ldots, k+T-1$,

$$S(\tau+1) = S(\tau) + \frac{1}{C_b}\left(\eta_{ch}E_{ch}(\tau) - \frac{E_{dsc}(\tau)}{\eta_{dsc}}\right) \quad (14)$$

$$0 \leq E_{ch}(\tau) \leq \delta_b(\tau)E_b^{nom} \quad (15)$$

$$0 \leq E_{dsc}(\tau) \leq (1-\delta_b(\tau))E_b^{nom} \quad (16)$$

$$S^{min} \leq S(\tau) \leq S^{max} \quad (17)$$

$$E_e(\tau) - E_{i,0}(\tau) = \hat{E}_{res}(\tau) - E_{ch}(\tau) + E_{dsc}(\tau) - \hat{E}_l(\tau) \quad (18)$$

$$0 \leq E_s(\tau) \leq E_e(\tau) \quad (19)$$

$$0 \leq E_s(\tau) \leq \hat{E}_c(\tau) \quad (20)$$

$$0 \leq E_{ch}(\tau) \leq \hat{E}_{res}(\tau) \quad (21)$$

$$0 \leq E_e(\tau) \leq E_g^{max}\delta_e(\tau) \quad (22)$$

$$0 \leq E_{i,0}(\tau) \leq E_e^{max}(1-\delta_e(\tau)) \quad (23)$$

where: $S$ is the State of Charge (SoC) of the BESS, which has a maximum $S^{max}$ and a minimum $S^{min}$ to be guaranteed, while $\eta_{ch}$, $\eta_{dsc}$, and $C_b$ are the BESS charge/efficiency efficiencies and capacity, respectively. $\hat{E}_{res}$, $\hat{E}_l$ and $\hat{E}_c$ are the forecasts of the RES generation, of the common internal load demand, and of the members aggregated energy demand, respectively.

At every time step $k$, we suppose to know the trajectories of forecasts $\hat{E}_{res}$, $\hat{E}_l$ and $\hat{E}_c$ and prices $p_e$ and $t_i$, over the optimization time horizon, and the current battery SoC $S(k)$.

In the optimization problem (13)-(23) we can observe that: constraints (14)-(16) implement the BESS SoC dynamics; constraint (18) imposes the REC energy balance; constraint (21) forces the battery to be charged only with renewable energy. Finally, in constraints (19) and (20), the shared energy $E_s$ is limited both by the exported energy $E_e$ and by the forecasted members aggregated energy demand $\hat{E}_c$. This indirectly imposes that $E_s = \min(E_e, E_c)$, according to (4). Indeed, in virtue of cost function (13) and of the definition of $I(k)$ in (1), $E_s$ will be maximized and thus leaded to be equal to the minimum among $E_e$ and $E_c$.

The considered study case is based on the data in [11]. It is composed by a set of 15 residential loads for a total nominal power of 45 kW. The PV generation reported in [11] has been scaled up to a nominal power of 50 kWp. TABLE I reports the other parameters of the study case. In the following we report the results obtained in 1 month of simulations. The zonal electricity price considered is referred to July 2021.

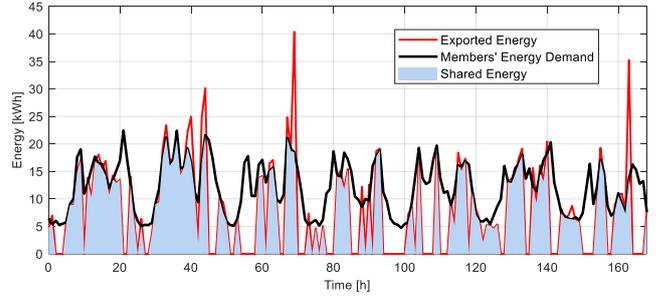

Fig. 6. First week of simulations: exported energy, aggregated energy demand of REC members and shared energy.

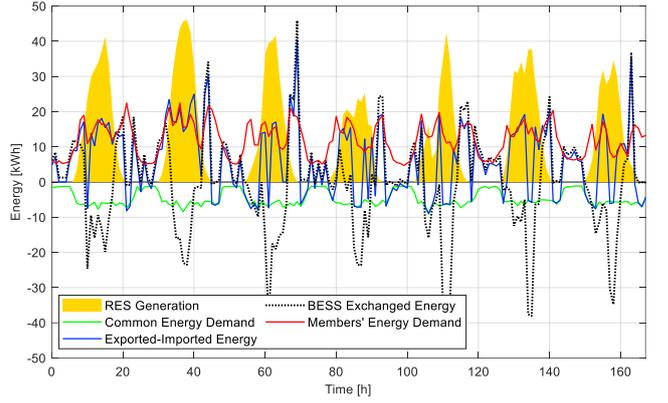

Fig. 7. First week of simulations: energy exchanges. Signs conventions: RES generation is posive; common energy demand is negative; exported-imported energy is positive when exporting; BESS exchanged energy is positive when discharging; memebers energy demand is positive.

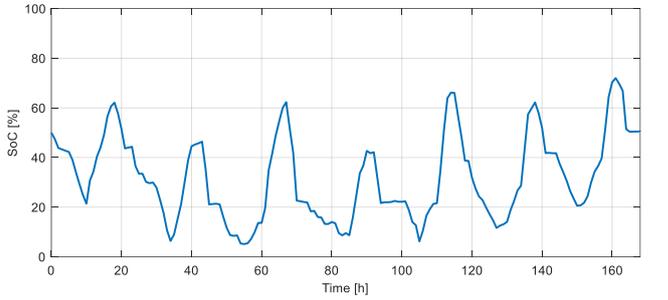

Fig. 8. First week of simulations: evolution of BESS SoC.

For the sake of readability, Figs. 6-8 show the details of the results obtained in the first week of simulations. In Fig. 6 we can observe how, as expected, the shared energy is always the minimum between the consumed energy by the REC members and the exported energy. Fig. 7 reports all the REC energy exchanges. Here we can observe that, when the PV generation (yellow area) is higher than the aggregated members demand (red line), a portion of the PV generation is used to recharge the battery (black line). The stored energy is then exported when PV generation is low (or zero during the night) but the aggregated members demand still is significant. In this way, the shared energy is maximized. Fig. 8 shows SoC of the BESS, which is always limited between 95% and 5%, as required.

The REC earning at the end of the simulated month accounts for 1063.43 €.

## VI. Conclusions

In this paper we presented the objectives, the organization and the first results of the ComER project, developed by the University of Cassino and the Campus Bio-Medico University of Rome. The main objective of the project is to develop methods and tools necessary for the management and control of RECs. In particular, the paper illustrated the rules stated by the Italian legislation for constituting and operating a REC, togheter with the incentives recognized to the REC members. Then three potential REC configurations are defined and associated to a proper mathematical formulation. Finally, the first results regarding the energy consumption forecasts and the optimal management of one of the configurations have been showed.